\documentclass[12pt]{iopart}
\usepackage{graphicx}
\usepackage{caption2}

\begin{document}

\title[]{Energy Dependence of High Moments for Net-proton Distributions}

\author{X.F. Luo$^{1,2}$\footnote{E-mail address: xfluo@lbl.gov}, B. Mohanty$^3$, H.G. Ritter$^1$, N. Xu$^1$
 }

\address{$^1$Lawrence Berkeley National Laboratory Berkeley, CA 94720, USA}
\address{$^2$Department of Modern Physics, University of Science and Technology of China, Heifei, Anhui 230026, China}
\address{$^3$Variable Energy Cyclotron Center Kolkata 700064, India}

\begin{abstract}High moments of multiplicity distributions of conserved quantities are predicted to
be sensitive to critical fluctuations.  To understand the effect of
the non-critical physics backgrounds on the proposed observable, we
have studied various moments of net-proton distributions with AMPT,
Hijing, Therminator and UrQMD models, in which no QCD critical point
physics is implemented. It is found that the centrality evolution of
various moments of net-proton distributions can be uniformly
described by a superposition of emission sources. In addition, in
the absence of critical phenomena, some moment products of
net-proton distributions, related to the baryon number
susceptibilities in Lattice QCD calculations, are predicted to be
constant as a function of the collision centrality. We argue that a
non-monotonic dependence of the moment products as a function of the
beam energy may be used to locate the QCD critical point.

\end{abstract}




\section{Introduction}
Heavy-ion reactions at high energy allow us to study the QCD phase
diagram experimentally~\cite{adams}. At vanishing baryon chemical
potential ($\mu_{B}=0$), Lattice QCD calculations predict that a
cross-over from the hadronic phase to the Quark Gluon Plasma (QGP)
phase will occur above a critical temperature. The temperature range
for the cross-over has been estimated to be about 170 - 190 MeV
\cite{aoki}. QCD based model calculations indicate that at large
$\mu_{B}$ the transition from the hadronic phase to the QGP phase
could be of first order with a critical point at the boundary to the
cross-over, the QCD Critical Point (QCP) \cite{ejiri}.  The location
of the QCP or even its existence are not confirmed ~\cite{misha}.
The possibility of the existence of the QCP has motivated our
interest to search for it with the RHIC beam energy scan
program~\cite{stephan}. By decreasing the collision energy down to a
center of mass energy of 5 GeV we will be able to vary the
baryo-chemical potential from $\mu_{B} \sim 0$ to $\mu_{B}$ of about
500 MeV.

A characteristic feature of a critical point is the increase and
divergence of the correlation length ( $\xi $ ) and of critical
fluctuations. In heavy-ion reactions, finite size effects, rapid
expansion, and critical slowing down could wash out those effects.
For example, the critical correlation length in heavy-ion collisions
is expected to be about 2-3 $fm$ \cite{stephanov}. A clear signature
of a critical point in an energy scan would be non-monotonic
behavior of non-gaussian multiplicity fluctuations.

Recently, theoretical calculations have shown that high moments of
multiplicity distributions of conserved quantities, such as
net-baryon, net-charge, and net-strangeness, are sensitive to the
correlation length $\xi$ \cite{stephanov_2}.

In Lattice QCD calculation with $\mu_{B}=0$, higher order
susceptibilities of the baryon number, which can be related to the
higher order moments of the net-baryon multiplicity distributions,
show a non-monotonic behavior near $T_{c}$ \cite{cheng}. A similar
behavior is expected for the finite $\mu_{B}$ region.
Experimentally, it is hard to measure the net-baryon number while
the net-proton number is measurable. Theoretical calculations show
that fluctuations of the net-proton number can be used to infer the
net-baryon number fluctuations at the QCP \cite{Hatta}.

In this paper we study the energy dependence of net-proton
multiplicity distributions for several models in terms of Skewness (
{\it S} ) and Kurtosis ( $\kappa$ ). This is a feasibility study for
the future data analysis from  the energy scan at RHIC.

\section{Observables}

We introduce various moment definitions of the event-by-event
multiplicity distributions: Mean, $M = $ $ <N>$, Variance,
$\sigma^{2}$ $=$ $ <(\Delta N)^{2}>$, Skewness, $S={{ < (\Delta N)^3
> }}/{{\sigma ^3 }}$, and Kurtosis, $\kappa={{ < (\Delta N)^4  > }}/{{\sigma ^4 }} - 3
$, where $\Delta N=N-<N>$. Skewness and Kurtosis are used to
characterize the asymmetry and peakness of the multiplicity
distributions, respectively. They are also used to demonstrate the
non-Gaussian fluctuation feature near the QCP, in particular a sign
change of the skewness may be a hint of crossing the phase boundary
\cite{stephanov_2,asakawa}.

To understand the centrality evolution of these moments, we
introduce the {\it Identical Independent Emission Source} ({\it
IIES}) assumption. Here the colliding system consists of a large
number of emission sources and the final multiplicity of particles
is the sum of the multiplicities from individual emission sources.
The relation between various moments and the number of emission
sources for the $i^{th}$ centrality can be expressed as:
$$(1): \frac{{M_i}}{{N_i }} =
\frac{{\sum\limits_{i = 1}^n {M_i } }}{{\sum\limits_{i = 1}^n {N_i
}}}  =  M(x),(2): \frac{{\sigma^2_i }}{{N_i }} =
\frac{{\sum\limits_{i = 1}^n {\sigma^2 _i}}}{{\sum\limits_{i = 1}^n
{ {N_i } } }} = \sigma^2(x)$$
$$(3): \frac{{N_i }}{{(1/S_i ^2 )}}
= \frac{{\sum\limits_{i = 1}^n {N_i}}}{{\sum\limits_{i = 1}^n
{(1/S_i ^2 )} }}  = S^2(x),(4): \frac{{N_i }}{{(1/\kappa _i )}} =
\frac{{\sum\limits_{i = 1}^n {N_i } }}{{\sum\limits_{i = 1}^n
{(1/\kappa _i)} }} = \kappa(x)$$

where $M_{i}, \sigma_{i},S_i,\kappa_i$ ($i=1,2,...n$) are the
moments extracted from the multiplicity distribution of the $i^{th}$
centrality and $N_i$ is the corresponding number of emission
sources. $M(x),\sigma(x), S(x) , \kappa(x)$ are various moments of
the multiplicity distributions for each emission source. From Equs
.(1)-(4), we obtain:
$$(5): \frac{{M_i }}{{\sum\limits_{i = 1}^n {M_i } }} = \frac{{\sigma _i ^2
}}{{\sum\limits_{i = 1}^n {\sigma _i ^2 } }} = \frac{{1/S_i ^2
}}{{\sum\limits_{i = 1}^n {(1/S_i ^2) } }} = \frac{{1/\kappa _i
}}{{\sum\limits_{i = 1}^n {(1/\kappa _i) } }} = \frac{{N_i
}}{{\sum\limits_{i = 1}^n {N_i } }}$$ which shows the connection
between the emission source distributions and the various moments of
multiplicity distributions. To investigate the centrality evolution
of those moments, we fit the normalized mean value ${{M_i
}}/{{\sum\limits_{i = 1}^n {M_i } }}$ with a function
$f(<N_{part}>)$, where $<N_{part}>$ is the average number of
participants. Then, we obtain:

\begin{eqnarray}
(6): M (<N_{part}> ) &=& \Big(\sum\limits_{i = 1}^n {M_i }\Big)
*f(<N_{part}>)  \nonumber \\ (7):\sigma (<N_{part}>) &=& \sqrt
{\Big(\sum\limits_{i = 1}^n {\sigma _i^2}\Big )*f(<N_{part}> )} \nonumber\\
(8): S (<N_{part}>  ) &=&
1\Bigg/\sqrt {\Big(\sum\limits_{i = 1}^n {1/S_i ^2 }\Big )*f(<N_{part}> )} \nonumber \\
(9):\kappa (<N_{part}> ) &=& 1\Big/[\Big(\sum\limits_{i = 1}^n
{1/\kappa _i }\Big )*f(<N_{part}>) ] \nonumber
\end{eqnarray}

Consequently, the centrality evolution of various moments can be
uniformly described by the function $f(<N_{part}>)$. From those
equations it also follows that the moment products, $S \sigma$,
$\kappa \sigma/S$ and $\kappa \sigma^2$, are constant as a function
of $<N_{part}>$.

\section{Results}

We calculated the various moments of net-proton ($\Delta
p=N_{p}-N_{\bar{p}}$) distributions from transport models ( AMPT
\cite{Lin}, Hijing \cite{XinNian}, UrQMD \cite{petersen}) and a
thermal model (Therminator \cite{Kisiel}). By using several models
with different physics implemented, we can study the effects of
physics correlations and backgrounds that are trivially present in
the data and that might modify purely statistical emission patterns,
like resonance decays, jet-production (Hijing), coalescence
mechanism of particle production (AMPT), thermal particle production
(Therminator), and hadronic rescatterring (AMPT,UrQMD).

\begin{figure}[htbp]
\centering \hspace{0cm} \vspace{-2ex}
\includegraphics[scale=0.4]{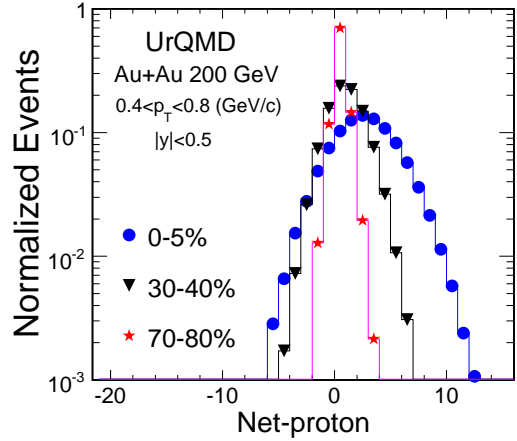}
\caption{\label{fig1} Typical event by event net-proton multiplicity
distributions of various centralities for Au+Au collisions at
$\sqrt{s_{NN}}=200$ GeV calculated by UrQMD model. }
\end{figure}

The kinetic coverage of protons and anti-protons used in our
analysis is $0.4<p_{T}<0.8$ GeV/c and $|y|<0.5$. In Fig.\ref{fig1},
typical net-proton distributions for three centralities,
$0-5\%,30-40\%$, and $70-80\%$, of Au+Au collisions at
$\sqrt{s_{NN}}=200$ GeV are calculated with the UrQMD model. The
shapes of net-proton distributions are different for the three
centralities. For the most central collisions,  $0-5\%$, the
net-proton distribution is wider compared to more peripheral
collisions. The discrepancies in shapes will be reflected in the
values of the different moments.

\begin{figure}[htbp]

\begin{minipage}[t]{0.5\linewidth}
\centering \vspace{0pt}
    \includegraphics[scale=0.4]{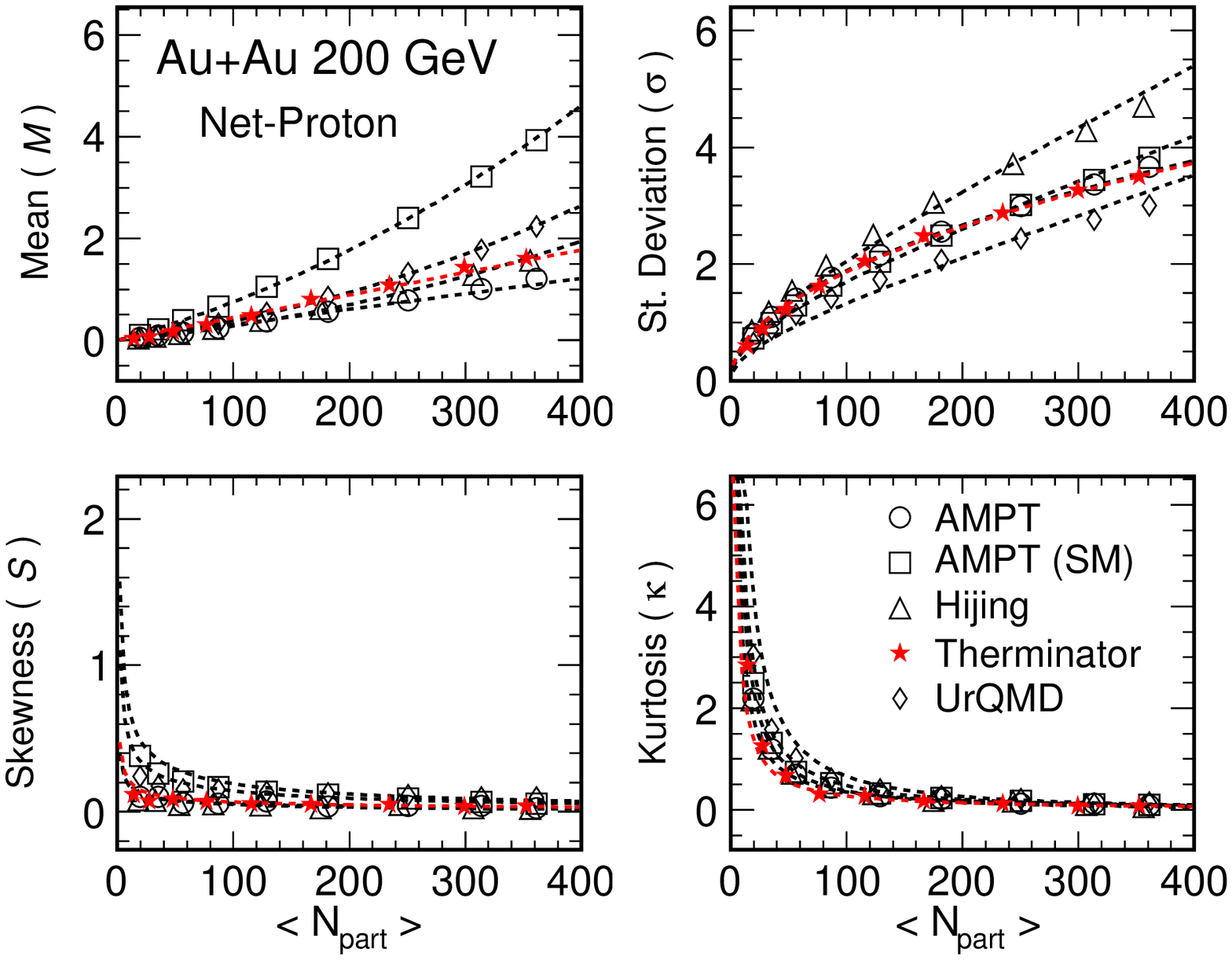}
\vspace{-0.3in}
   \caption{Centrality dependence of various moments of $\Delta p$ distributions for Au+Au collisions at $\sqrt{s_{NN}}=200$ GeV from various models. The dashed lines
   represent the expectations for statistical emission. }
    \label{fig:side:a}
  \end{minipage}%
  \hspace{0.2in}
  \begin{minipage}[t]{0.5\linewidth}
  \centering \vspace{0pt}
   \includegraphics[scale=0.4]{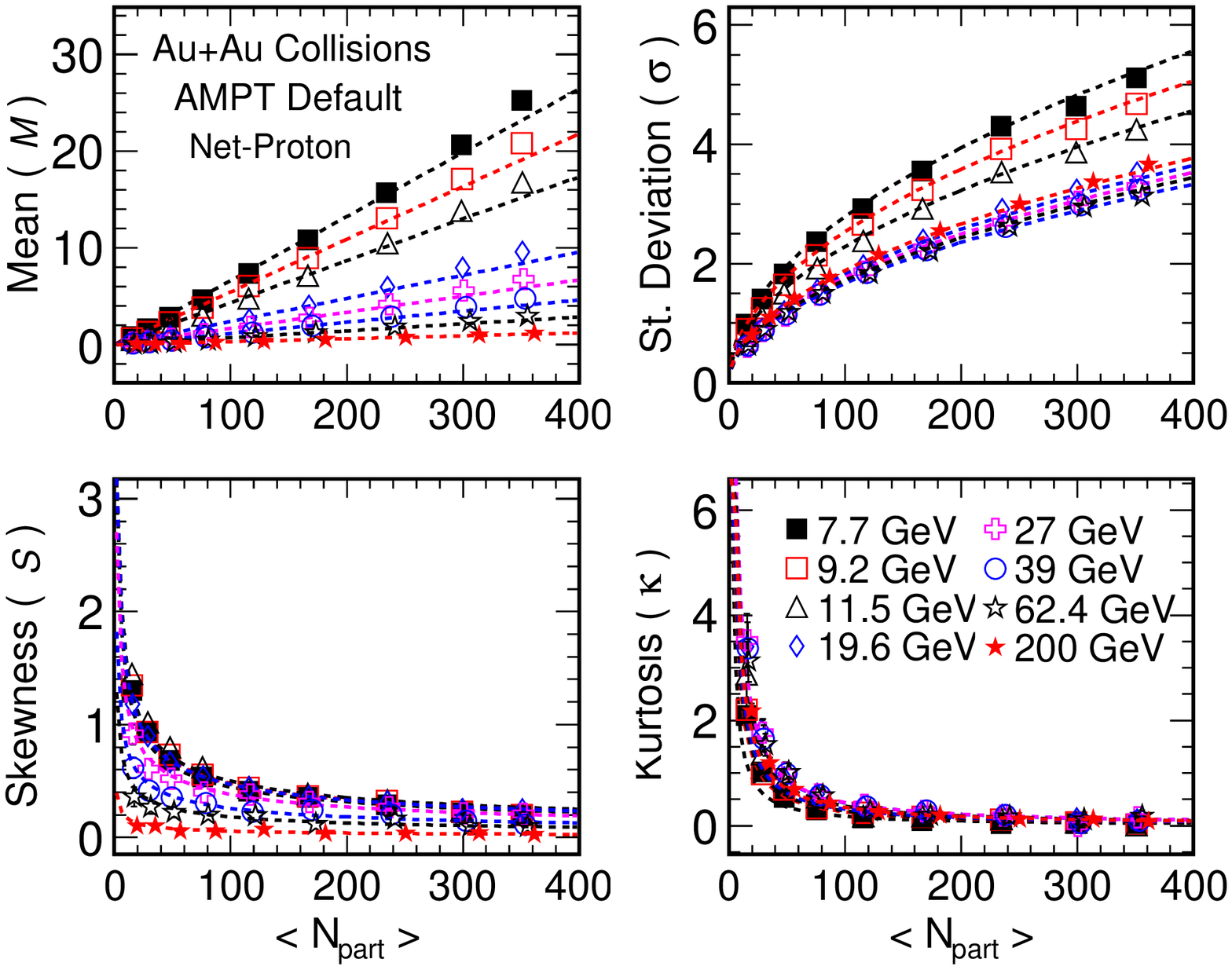}
     \vspace{-0.3in}
    \caption{Centrality dependence of various moments of $\Delta p$  distributions for Au+Au collisions at various energies from the AMPT model. The dashed lines
   represent the expectations for statistical emission.}
    \label{fig:side:b}
  \end{minipage}

\end{figure}

Fig. 2 shows the $< N_{part }
>$ dependence of four moments ({\it M}, $\sigma$, {\it S}, $\kappa$)
extracted from net-proton distributions of Au+Au collisions at
$\sqrt{s_{NN}}=200$ GeV for the various models. {\it M} and $\sigma$
show a monotonic increase with $<N_{part}>$ for all of the models,
while {\it S} and $\kappa$ decrease monotonically. In Fig. 3, we
choose the default AMPT model to evaluate the centrality evolution
of the various moments of net-proton distributions for various
energies. {\it M} shows a linear increase with $<N_{part}>$ and a
decrease with $\sqrt{s_{NN}}$. $\sigma$ increases monotonically with
$<N_{part}>$ while it has non-monotonic dependence on
$\sqrt{s_{NN}}$. {\it S} is positive and decreases with increasing
$<N_{part}>$ and $\sqrt{s_{NN}}$. The net-proton distributions
become more symmetric for central collision and higher energies.
$\kappa$ decreases with $<N_{part}>$ and is similar for all
energies. The dashed lines in Fig. 2 and Fig. 3 are resulting from
Equs. (6)-(9) to evaluate the centrality evolution of the various
moments. To apply our formulas, we fit the normalized mean value in
Equ.(5) with the function $f(<N_{part}>)$. For AMPT String Melting
(SM) \cite{Lin}, Hijing \cite{XinNian} and UrQMD \cite{petersen}
models, a 2nd order polynomial,
$f(<N_{part}>)=a<N_{part}>^2+b<N_{part}>$ is applied, while a linear
function $f(<N_{part}>)=a*<N_{part}>$ is employed for AMPT default
\cite{Lin} and Therminator \cite{Kisiel} models. Once the function
$f(<N_{part}>)$ is obtained, the centrality evolution of the other
moments is completely determined by Equs. (7)-(9). It is obvious
that the centrality evolution of the various moments of net-proton
distributions  in Fig. 2 and in Fig. 3 can be well described by the
dashed lines.

\begin{figure}[htbp]

\begin{minipage}[t]{0.5\linewidth}
\centering \vspace{0pt}
  \includegraphics[scale=0.4]{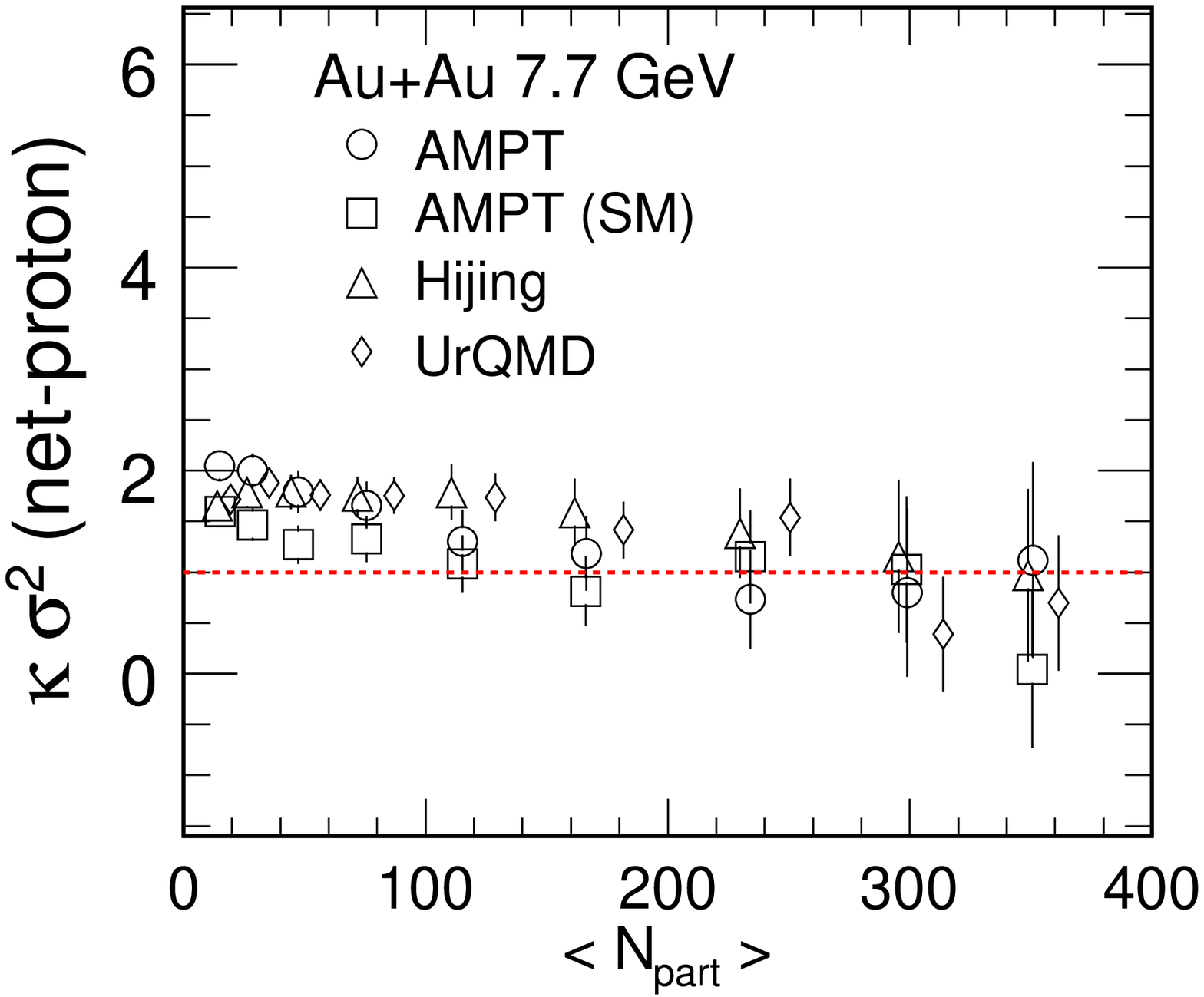}
   \vspace{-0.3in} \caption{The $\kappa \sigma^{2}$ of net-proton  distributions for
   Au+Au 7.7 GeV collisions as a function of $<N_{part}>$ from various
    models.}
  \label{fig4}
  \end{minipage}%
  \hspace{0.15in}
  \begin{minipage}[t]{0.5\linewidth}
  \centering \vspace{0pt}
   \includegraphics[scale=0.4]{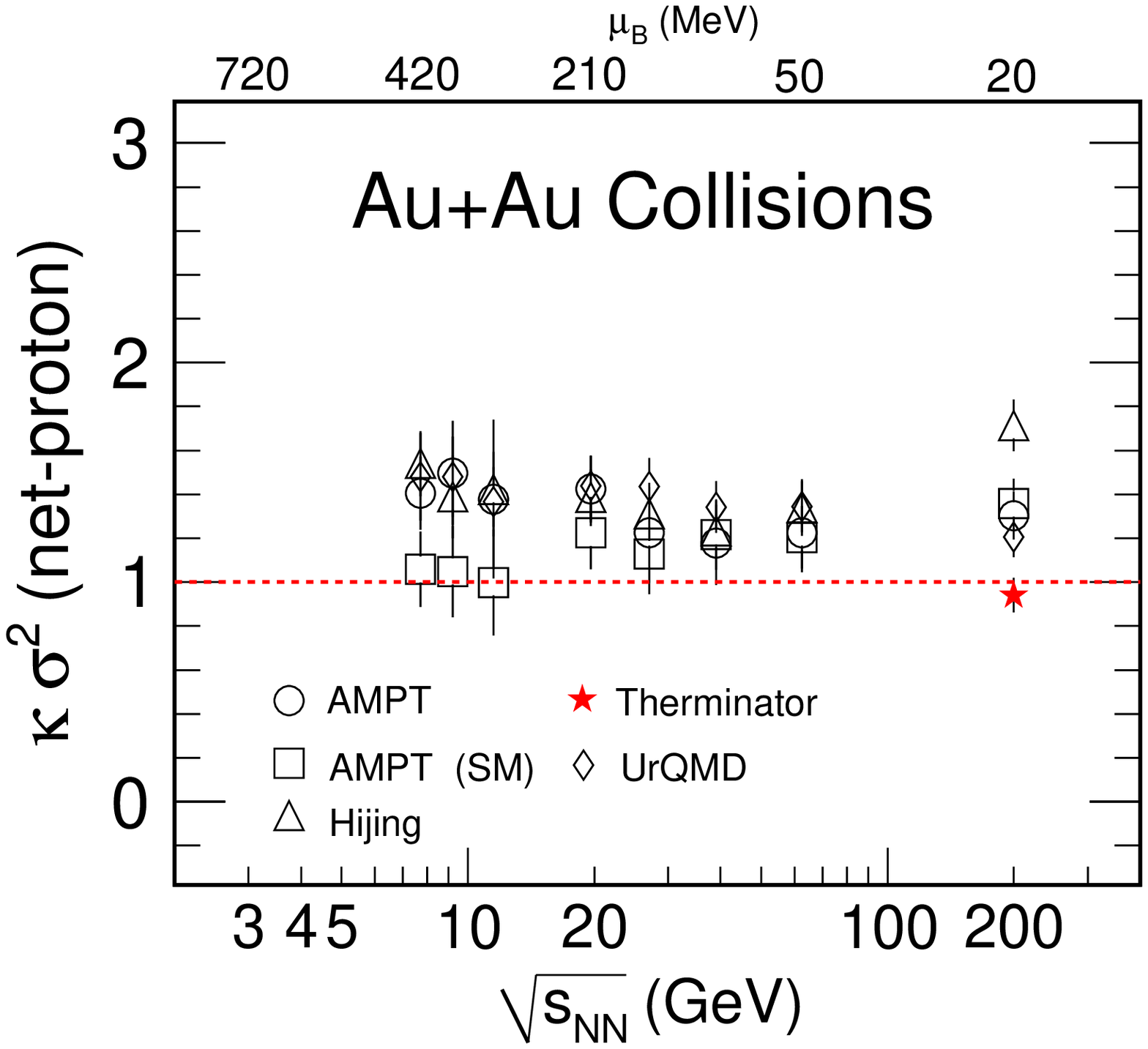}
 \vspace{-0.3in} \caption{The $\kappa \sigma^{2}$ of net-proton distributions in Au+Au collisions as a function of $\sqrt{s_{NN}}$ for various models. }
    \label{fig5}

  \end{minipage}
\end{figure}

The $\kappa\sigma^2$ of the net-proton distributions of Au+Au
collisions at $\sqrt{s_{NN}}=7.7$ GeV as a function of $<N_{part}>$
is shown in Fig. 4 for various models. $\kappa\sigma^2$ is constant
with respect to $<N_{part}>$ within the errors, which is consistent
with the expectation from the {\it IIES} assumption. Fig. 5 shows
the energy dependence of the $\kappa \sigma^2$ of net-proton
distributions for various models.  The top of the figure shows the
$\mu_{B}$ value corresponding to the various energies. The values
shown are averaged within the centrality range studied. The results
from various models show no dependence on energy and are close to
unity. This suggests that the $\kappa \sigma^2$ of net-proton
distributions is not affected very much by the non-QCP physics at
different beam energies, such as the change of $\mu_B$ \cite{adams},
and the collective expansion \cite{flow}. Note that the result from
the pure thermal model, Therminator, is much closer to unity
compared to others. Actually, if proton and anti-proton have
independent poisson distributions, the difference of protons and
anti-protons should distribute as a Skellam distribution
\cite{skellam}, for which $\kappa \sigma^2$ is unity. A large
deviation from constant as a function of $<N_{part}>$ and collision
energy for $\kappa \sigma^2$ may indicate new physics, such as
critical fluctuations.

\section{Summary and Outlook}

Higher moments of the distribution of conserved quantities are
predicted to be sensitive to the correlation length at QCP and to be
related to the susceptibilities computed in Lattice QCD. Various
non-QCP models (AMPT, Hijing, Therminator, UrQMD) have been applied
to study the non-QCP physics background effects on the high moments
of net-proton distributions. The centrality evolution of the high
moments from models can be well described by the scaling derived
from the {\it IIES} assumption and  the moment products $S \sigma$,
$\kappa \sigma/S$  and $\kappa \sigma^2$ of net-proton distributions
are constant with respect to $<N_{part}>$. $\kappa \sigma^2$ is also
found to be constant as a function of energy for various models.

Our model study can serve as a background study of the behavior
expected from known physics effects for the RHIC beam energy scan,
that will span values of $\mu_{B}$ from 100 to about 550 MeV. The
presence of a critical point in that region may result in
non-gaussian fluctuations and in correlated emission. Then the {\it
IIES} assumption will break down. This is expected to lead to
non-monotonic behavior of the observables studied here as a function
of collision energy.

\section*{Acknowledgments}
This work was supported in part by the U.S. Department of Energy
under Contract No. DE-AC03-76SF00098 and National Natural Foundation
of China under Grant No. 10835005. BM thanks the Department of
Atomic Energy, Government of India for financial support.

\section*{References}

\bibliography{SQM2009}
\bibliographystyle{unsrt}

\end{document}